\documentstyle[psfig]{kapproc} 

\let\footnote\savefootnote

\kluwerbib

\newcommand \kms{km s$^{-1}$~}

\newcommand \msunyr{M$_\odot$yr$^{-1}$}

\begin{document}



\articletitle[]{Magnetically channeled accretion in T Tauri stars : a
dynamical process}


\author{J. Bouvier$^1$, S.H.P. Alencar$^2$, C. Dougados$^1$}
\affil{$^1$ Observatoire de Grenoble, $^2$ UFMG Belo Horizonte}



 \begin{abstract}
   We review observational evidence and open issues related to the process
   of magnetospheric accretion in T Tauri stars. Emphasis is put on recent
   numerical simulations and observational results which suggest that the
   interaction between the stellar magnetosphere and the inner accretion
   disk is a highly time dependent process on timescales ranging from hours
   to months.
 \end{abstract}


\section{Introduction}

T Tauri stars are low-mass stars with an age of a few million years at
most, still contracting down their Hayashi tracks towards the main
sequence. Many of them, the so-called classical T Tauri stars, show
signs of accretion from a circumstellar disk (see, e.g., M\'enard \&
Bertout 1999 for a review). Understanding the accretion process in
young solar type stars is one of the major challenges in the study of
T Tauri stars.  Indeed, accretion has a significant and long lasting
impact on the evolution of low mass stars by providing both mass and
angular momentum while the evolution and ultimate fate of
circumstellar accretion disks have become an increasingly important
issue since the discovery of extrasolar planets and planetary systems
with unexpected properties.  Deriving the properties of young stellar
systems, of their associated disks and outflows is therefore an
important step towards the establishment of plausible scenarios for
star and planet formation.

\begin{figure}[ht]
\centerline{\psfig{file=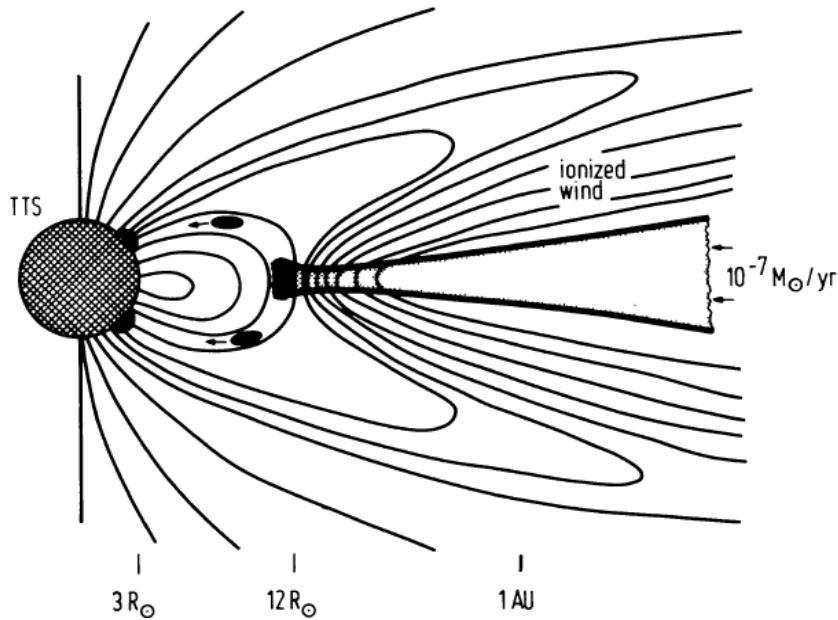,width=\textwidth}}
\caption{A sketch illustrating the basic concept of magnetospheric
        accretion in T Tauri stars (from Camenzind 1990).}
\label{cam}
\end{figure}

Strong surface magnetic fields have long been suspected to exist in
TTS based on their powerful X-ray and centrimetric radio emissions
(Montmerle et al. 1983, Andr\'e 1987). Surface fields of order of 1-3
kG have recently been derived from Zeeman broadening measurements of
CTTS photospheric lines (Johns Krull et al. 1999, 2001; Guenther et
al. 1999) and from the detection of electron cyclotron maser emission
(Smith et al. 2003). These strong stellar magnetic fields are believed
to significantly alter the accretion flow in the circumstellar disk
close to the central star. Based on models originally developped for
magnetized compact objects in cataclysmic binaries (Ghosh \& Lamb
1979) and {\it assuming\/} that T Tauri magnetospheres are
predominantly dipolar on the large scale, Camenzind (1990) and
K\"onigl (1991) showed that the inner accretion disk is expected to be
truncated by the magnetosphere at a distance of a few stellar radii
above the stellar surface for typical mass accretion rates of
10$^{-9}$ to 10$^{-7}$\msunyr\ in the disk (Basri \& Bertout 1989;
Hartigan et al. 1995; Gullbring et al. 1998). Disk material is then
channeled from the disk inner edge onto the star along the magnetic
field lines, thus giving rise to magnetospheric accretion columns. As
the free falling material in the funnel flow eventually hits the
stellar surface, accretion shocks develop near the magnetic poles. The
basic concept of magnetospheric accretion in T Tauri star is
illustrated in Figure~\ref{cam}. The successes and limits of current
magnetospheric accretion models (MAMs) in accounting for the observed
properties of classical T Tauri systems (CTTS) are reviewed in Section
2, while in Section 3 we discuss recent results which suggest that the
interaction between the star's magnetosphere and the inner disk is a
highly dynamical and time dependent process.

\section{Magnetospheric accretion : the static view}

The magnetospheric accretion scenario makes clear predictions regarding the
existence of an inner (magnetospheric) cavity with a radial extent of a few
stellar radii, magnetic accretion columns filled with free falling plasma,
and localized accretion shocks at the surface of CTTS. Observational
support for these various predictions is reviewed in this section.

\subsection {Magnetospheric cavity and IR excess} 

Disk truncation radii of typically 3-8R$_\star$, as predicted by MAMs, are
required to reproduce the near-IR spectral energy distribution of some CTTS
(Bertout et al. 1988) and to account for the properties of CTTS infrared
excess on a statistical basis (Meyer et al. 1997). However, recent
near-infrared monitoring studies of accreting T Tauri stars have revealed
large variability of some systems occurring on short timescales at these
wavelengths (Eiroa et al. 2002) which are not easily accounted for by
static MAMs and probably points to time variable accretion in the inner
disk. Moreover, the near-IR veiling measured in classical T Tauri systems
is often larger than predicted by standard disk models (Folha \& Emerson
1999, Johns-Krull \& Valenti 2001) which suggests that the inner disk
structure is significantly affected by its interaction with the stellar
magnetosphere and departs from flat disk models. Observational evidence for
an inflated inner disk edge has been reported by Bouvier et al.  (1999) and
is thought to result from the interaction between the disk and an {\it
  inclined\/} stellar magnetosphere (Lai 1999, Terquem \& Papaloizou 2000).

\subsection {Accretion columns and line profiles} 

The common occurrence of inverse P Cygni profiles in the higher Balmer
lines of CTTS spectra, with redshifted absorption components reaching
velocities of several hundred \kms, points to gas being accreted onto the
star at free fall velocity from a distance of several stellar radii
(Edwards et al. 1994). The relative success of MAMs in reproducing the
shape and strength of optical and near IR emission line profiles of some
CTTS provides additional support to the formation of at least part of the
line emission in accretion columns (Hartmann et al. 1994; Muzerolle et al.
2001; for a more critical view see Ardila et al. 2002b, Alencar \& Basri
2000, Folha \& Emerson 2001).  Statistical correlations between line fluxes
and mass accretion rates predicted by these models have been reported in
samples of TTS for a variety of emission lines in the UV, optical, and near
IR range (e.g.  Beristain et al. 2001, Johns-Krull et al.  2000, Alencar \&
Basri 2000, Folha \& Emerson 2001, Muzerolle et al. 2001).  Correlations
have also been reported between mass loss rate and mass accretion rate
(e.g., Ardila et al. 2002) as expected from MAMs which predict that part of
the accretion flow is diverted into the wind by the magnetic field (e.g.
Shu et al. 1994, Ferreira 1997).

In a few CTTS, synoptic studies have revealed periodic variations of the
line profiles intensity and shape over a rotation timescale. The rotational
modulation of line flux and radial velocity has been interpreted as
resulting from the changing visibility and projected geometry of an {\it
  inclined\/} magnetosphere interacting with the disk (Johns \& Basri 1995,
Petrov et al. 1996, 2001, Oliveira et al. 2000, Batalha et al. 2002).
Finally, two Doppler imaging studies of CTTS (Unruh et al. 1998,
Johns-Krull \& Hatzes 1997) have reported the existence of localized spots
at the surface of CTTS whose maximum visibility was found to coincide with
the developement of high velocity redshifted absorptions in emission line
profiles, as expected from MAMs when the line of sight to the accretion
shock intersects the accretion column.

All these observational results thus tend to support the existence of ({\it
  tilted}) magnetospheric accretion columns in CTTS. To be fair, however,
one must stress that the modulation of inflow and outflow signatures in
line profiles has been reported so far for only a few (well studied)
systems. Moreover, even in these cases, multiple periods are sometimes
observed and their relationship to the stellar rotation period is not
always clear (e.g. Alencar \& Batalha 2002, Oliveira et al. 2000).  In
addition, some of the basic predictions of MAMs, e.g. correlated time
variations of line flux (from the accretion columns) and continuum excess
flux (from the accretion shocks) are not always observed (Ardila \& Basri
2000, Batalha et al. 2002). And some of the statistical correlations
expected from MAMs for CTTS samples are not always fulfilled (Johns-Krull
\& Gafford 2002).

Hence, while several key properties of CTTS emission line profiles are
consistent with MAMs predictions and a few systems indeed appear to behave
as qualitatively expected from these models, additional processes are
likely present which blur the clear spectral signatures of steady state
magnetospheric accretion and associated mass loss. The occasional failure
to observe some of the basic predictions of static MAMs may result at least
in part from the intrinsically unstable nature of the magnetospheric
accretion process as we discuss in the section 3.

\subsection {Accretion shocks, UV lines and veiling}
   
Further observational support for MAMs comes from the detection of the
rotational modulation of CTTS luminosity by bright surface spots. Modelling
the periodic light variations observed on a timescale of the stellar
rotation period suggests hot spots covering of order of one percent of the
stellar surface. The localized hot spots are thus identified with the
accretion shocks expected to develop near the magnetic poles at the base of
magnetic accretion columns (Bouvier \& Bertout 1989; Vrba et al.  1993).
Continuum flux excesses from the UV to the optical range are expected to be
associated with the accretion shock and, at least in some cases, correlated
variations between the system's brightness and veiling or excess flux
supports this expectation (e.g. Chelli et al. 1999).
   
Accretion shock models predict the formation of high excitation lines in
the shock region (Calvet \& Gullbring 1998, Lamzin 1995) and empirical
evidence is consistent with the formation of narrow line profiles in the
postshock gas (e.g. HeI, Beristain et al. 2001) while broader profiles
would arise from the preshock region (e.g. Gomez de Castro \& Lamzin 1999).
However, the width, strength and kinematics of high excitation UV lines
have so far eluded model predictions being broader ($\sim$200 km/s, Ardila
et al.  2002a), stronger (Lamzin \& Kravtsova, this volume) and exhibiting
lower centroid velocities (Ardila \& Johns-Krull, this volume) than
expected from models.
   
Solving the discrepancies between MAMs predictions and observed line
profiles may require the inclusion of a turbulent component in the
accretion flow (Ardila et al. 2002b, Alencar \& Basri 2000). Turbulence may
additionally provide a heating source for the plasma in the funnel flow, up
to values deduced from the observations ($\leq$10$^4$K) which non turbulent
models hardly reach (e.g. Martin 1996). Thus, turbulence could conceivably
generate magnetic waves whose damping in the accretion column would
contribute to the heating of the plasma (see Vasconcelos et al., this
volume). Clearly, improved shock models are required to yield a
self-consistent description of the physical conditions in the magnetic
accretion region of CTTS.

\subsection {Magnetic braking and angular momentum regulation}

T Tauri stars have surprisingly low rotation rates considering that they
accrete high angular momentum material from their disk. In the
magnetospheric accretion framework, the lever arm provided by extended
magnetic field lines threading the disk outside the corotation
radius\footnote{The corotation radius is the radius in the disk where the
  keplerian angular velocity of the disk material equals the angular
  velocity of the star. It is of order of a few stellar radii in TTS.}
allows the star to tranfer angular momentum outwards with a net braking
effect (Collier Cameron \& Campbell 1993; Armitage \& Clarke 1996; Shu et
al.  1994). This possibly accounts for the slower rotation rates of
accreting T Tauri stars compared to non accreting ones, as originally
reported by Bouvier et al. (1993) and Edwards et al. (1993). Recent and
more complete studies of TTS rotation rates have however failed to reveal
the expected correlation between low rotation and disk accretion, at least
for very low mass T Tauri stars (see Stassun et al. 1999; Herbst et al.
2000). Nevertheless, evidence has been accumulating that low mass pre-main
sequence stars evolve down their Hayashi tracks without accelerating
(Rebull et al. 2002), thus confirming the apparent paradox that the central
star loses large amounts of angular momentum during the accretion phase.
   
\section{Magnetospheric accretion : a dynamical process} 

Many of the expected signatures of magnetically mediated accretion have
thus been observed in CTTS. Evidence, however, comes mostly from snapshot
studies and it is only recently that the stability of the phenomenon has
started to be addressed. Time dependent modelling of the interaction
between the inner disk and the stellar magnetosphere requires heavy
numerical simulations and observational clues can only be obtained through
long term monitoring studies combining high resolution spectroscopy and
multi band photometry. In this section, we briefly review model predictions
regarding the temporal evolution of the process and recent observational
results which seem to indicate that it is highly time dependent indeed.

\subsection{Time dependent models} 

Most MAMs assume that the stellar magnetosphere truncates the disk close to
the corotation radius, i.e., the radius in the disk where the keplerian
angular velocity of the disk material equals that of the star. Field lines
threading the disk at this radius thus corotate with the central object.
However, due to the finite radial distance over which the stellar
magnetosphere interacts with the inner disk, the footpoints of most field
lines rotate differentially, one being anchored into the star, the other
into the keplerian disk.

Recent numerical simulations indicate that magnetic field lines are thus
substantially distorted by differential rotation on a timescale of only a
few keplerian periods at the inner disk. One class of models predict that
differential rotation leads to the expansion of the field lines, their
opening and eventually their reconnection which restores the initial
(assumed dipolar) magnetospheric configuration (e.g. Aly \& Kuijpers 1990;
Goodson et al.  1997; Goodson \& Winglee 1999). Magnetospheric {\it
  inflation cycles\/} are thus expected to develop and to be accompanied by
violent episodic outflows as field lines open and reconnect as well as time
dependent accretion rate onto the star (Hayashi et al.  1996; Romanova et
al. 2002).  The most recent 3D MHD simulations of disk accretion onto an
inclined stellar magnetosphere are presented in Romanova et al.  (2003) and
illustrate well the extreme complexity of the process.

Other models, however, suggest that the field lines respond to differential
rotation by drifting radially outwards in the disk, leading to magnetic
flux expulsion (Bardou \& Heyvaerts 1996).  The response of the magnetic
configuration to differential rotation mainly depends upon the magnitude of
magnetic diffusivity in the disk, a parameter of the models which is poorly
constrained from basic principles.

\subsection{Preliminary observational evidence} 

Due mostly to the lack of intense monitoring of CTTS on proper timescales,
the observational evidence for a time dependent interaction between the
inner disk and the stellar magnetosphere is at present limited. Episodic
high velocity outbursts, possibly connected with magnetospheric
reconnection events, have been reported for a few systems based on the
slowly varying velocity of blueshifted absorption components in emission
line profiles on a timescale of hours to days (Alencar et al.  2001; Ardila
et al. 2002a). Possible evidence for magnetic field lines being twisted by
differential rotation has been reported for SU Aur by Oliveira et al.
(2000). These authors measured a time delay of a few hours between the
appearance of high velocity redshifted absorption components in line
profiles formed at different altitudes in the accretion column. They
interpreted this result as the crossing of an azimuthally twisted accretion
column on the line of sight. Another possible evidence for magnetic field
lines being twisted by differential rotation and leading to quasi-periodic
reconnection processes has been reported for the embedded protostellar
source YLW 15 based on the observations of quasi-periodic X-ray flaring
(Montmerle et al.  2000).

\subsection{ AA Tau : the prototype of time variable magnetospheric accretion ?}

Large scale synoptic studies of a handful of CTTS have been performed in
the last years and provided new insight into the magnetospheric accretion
process and its temporal evolution. Two of these have targetted the CTTS AA
Tau which proved to be ideally suited to probe the inner few 0.01 AU of the
disk-magnetosphere interaction region. Due to its high inclination
($i\simeq$75$^o$, see Bouvier et al.  1999), the line of sight to the star
intersects the region where the stellar magnetosphere threads the disk. The
peculiar orientation of this otherwise typical CTTS maximizes the
variability induced by the modulation of the magnetospheric structure and
thus provides the strongest constraints on the inner disk and the
magnetospheric cavity.

A first monitoring campaign, whose results are reported in Bouvier et al.
(1999), led to the discovery of recurrent eclipses of the central object
with a period of 8.2 days. The eclipses were attributed to a non
axisymmetric warp of the inner disk edge which periodically obscures the
the central star as it orbits it at keplerian speed. Such an inner disk
warp is expected to develop as the disk encounters an {\it inclined}
magnetosphere (Terquem \& Papaloizou 2000; Lai 1999; Romanova et al.
2003).

A second campaign which combined photometry and high resolution
spectroscopy and whose results are reported in Bouvier et al. (2003) was
performed in 1999. The spectroscopic variability provided evidence for
accretion columns and associated hot spots with signatures (redshifted
absorptions, continuum excesses) modulated on the rotation timescale of the
system. In addition, a time delay of about 1 day was reported between the
flux variations of lines forming at a different altitude in the accretion
column, from H$_\alpha$ near the disk inner edge to HeI close to the
stellar surface. The measured time delay is consistent with accreted gas
blobs propagating downwards along the accretion column at free fall
velocity from a distance of about 8 R$_\star$, the radius at which the
stellar magnetosphere disrupts the inner disk. These line flux variations
indicate non steady accretion along the magnetic funnel flows onto the star
on short timescales.
   
On longer timescales, of order of one month, line and continuum excess flux
variations were also observed, indicative of a smoothly varying mass
accretion rate onto the star.  Simultaneously, a tight correlation was
observed between the radial velocity of the blueshifted and redshifted
absorption components in the H$_\alpha$ emission line profile (see Fig.2).
Since the former is a wind diagnostics while the latter forms in the
accretion flow, this correlation provides additional evidence for a
physical connection between (time dependent) inflow and outflow in CTTS.
Bouvier et al. (2003) argued that the associated flux and radial velocity
variations can be consistently interpreted in the framework of {\it
  magnetospheric inflation cycles}, as predicted by recent numerical
simulations of the star-disk interaction.

\begin{figure}[ht]
\sidebyside
{\centerline{\psfig{file=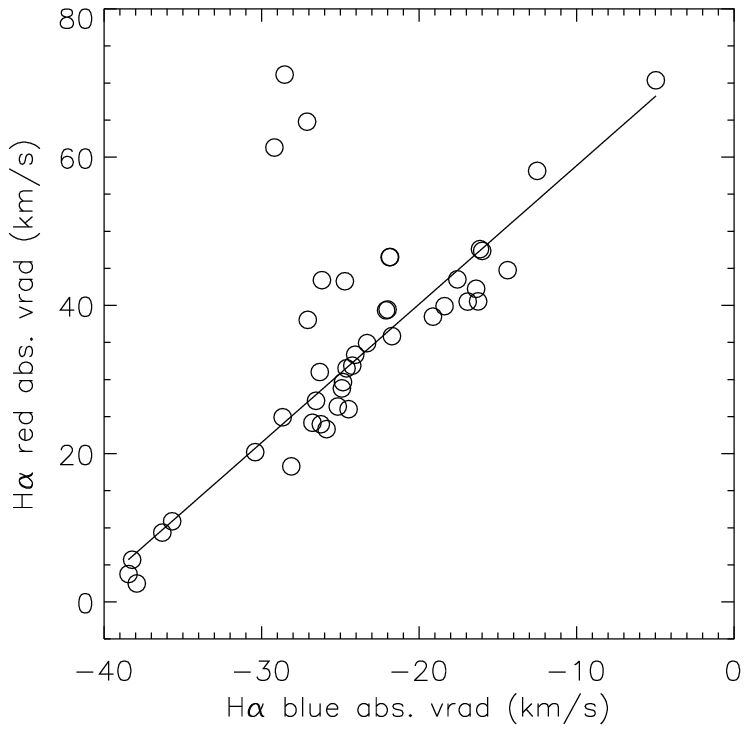,height=2.2in}}
\caption{The correlation between the radial velocity of the blueshifted
     and redshifted absorption components in the H$_\alpha$ emission line
     profile of AA Tau (from Bouvier et al. 2003).}}
{\centerline{\psfig{angle=270,file=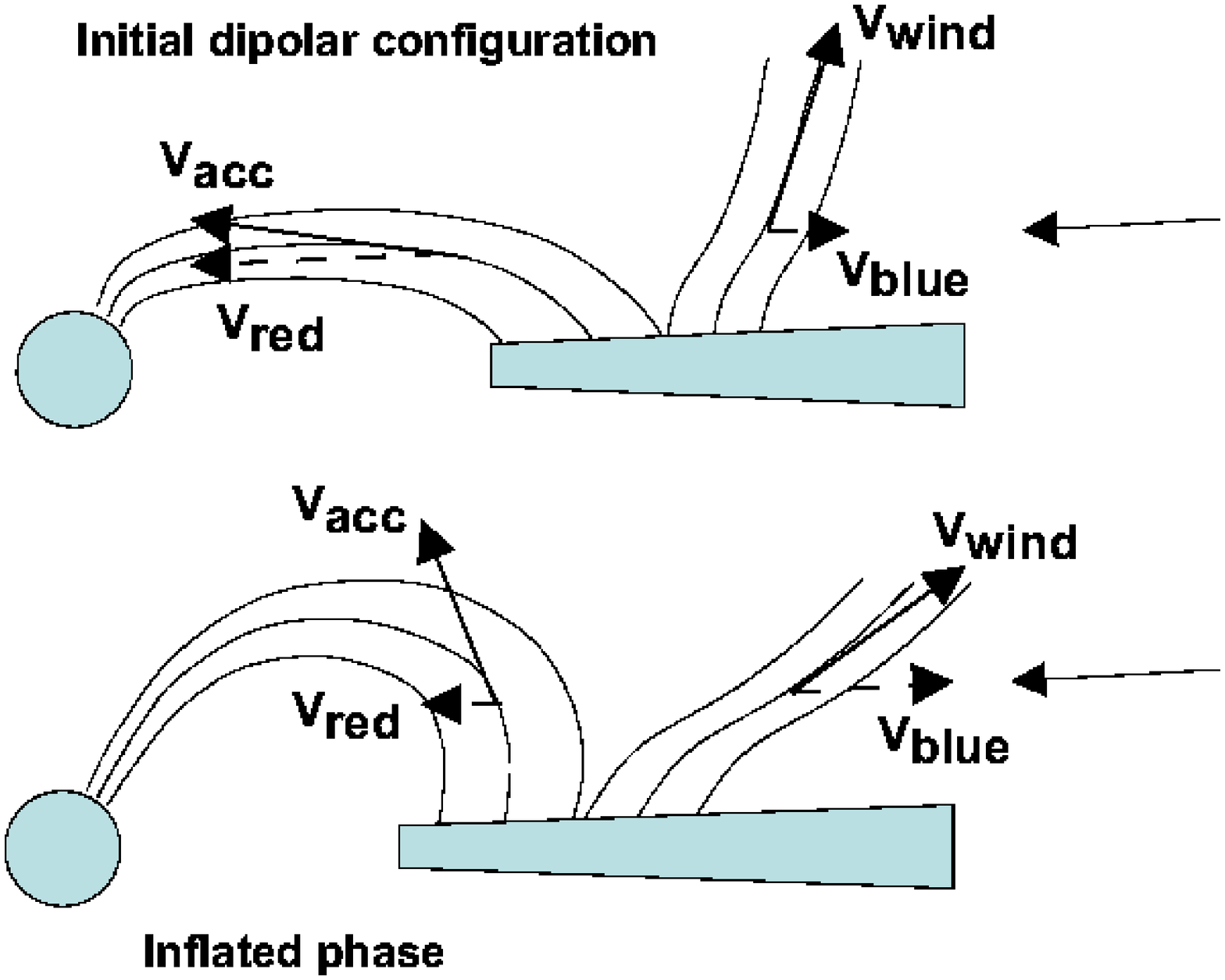,height=2.2in}}
\caption{A sketch of the magnetospheric inflation
  scenario. The arrow on the right side indicates the line of sight to the
  AA Tau system (from Bouvier et al. 2003).}}
\end{figure}
\noindent

This is schematically illustrated in Figure~3. As magnetic field lines
expand due to differential rotation between the star and the inner disk,
the radial velocity of the accretion (resp. wind) flow decreases (resp.
increases) due to projection effects on the line of sight. At the same
time, the loading of disk material onto inflated field lines becomes
increasingly difficult owing to the large angle field lines make relative
to the disk plane. This results in a reduced accretion rate onto the star,
as deduced from the depressed line and continuum fluxes observed at this
phase.

The last synoptic campaign on AA Tau thus yields the first evidence for
global instabilities developping on a timescale of a month in the large
scale structure of the magnetosphere, a result which provides support to
the predictions of time dependent models of magnetospheric accretion.
Whether these magnetospheric instabilities are cyclic, being driven by
differential rotation as predicted by numerical models, will require
additional monitoring lasting for several months.

\section{Conclusion}

Observational evidence for magnetospheric accretion being instrumental in
classical T Tauri stars has accumulated in recent years. Several key
properties of these young stars are naturally accounted for by assuming
that the stellar magnetic field governs the accretion flow close to the
star. The strong variability of CTTS on all timescales, from hours to
months (and possibly years, Bertout 2000), further suggests that the
magnetically mediated interaction between the accretion disk and the
central object is a highly dynamical and time dependent process.

The implications of the non steadiness of magnetospheric accretion in CTTS
are plentiful and remain to be fully explored. They range from the
evolution of their angular momentum (Agapitou \& Papaloizou 2000), the
origin of inflow/ouflow short term variability (Woitas et al. 2002,
Lopez-Martin et al. 2003), the modelling of the near infrared excess of
CTTS and of its variations both of which will be affected by a non standard
and time dependent inner disk structure (Carpenter et al. 2001, Eiroa et
al. 2002, Johns-Krull \& Valenti 2003), the origin of CTTS variability
which is expected to be a complex combination of modulation by hot and cold
spots and variable circumstellar extinction (e.g. DeWarf et al. 2003), and
possibly the halting of planet migration close to the star (Lin et al.
1996).

\begin{acknowledgments}
We thank the organisers for a very fruitful and enjoyable meeting, as well
as for financial support. 
\end{acknowledgments}


\begin{chapthebibliography}{1}

\bibitem[Agapitou \& Papaloizou(2000)]{2000MNRAS.317..273A} Agapitou, V.~\& 
Papaloizou, J.~C.~B.\ 2000, MNRAS, 317, 273 

\bibitem[2000]{ale00} Alencar, S.H.P., \& Basri, G. 2000, AJ, 119, 1881

\bibitem[Alencar \& Batalha(2002)]{2002ApJ...571..378A} Alencar, 
S.~H.~P.~\& Batalha, C.\ 2002, ApJ, 571, 378 

\bibitem[Alencar, Johns-Krull, \& Basri(2001)]{2001AJ....122.3335A} 
Alencar, S.~H.~P., Johns-Krull, C.~M., \& Basri, G.\ 2001, AJ, 122, 3335 

\bibitem[Aly \& Kuijpers(1990)]{1990A&A...227..473A} Aly, J.~J.~\& 
Kuijpers, J.\ 1990, A\&A, 227, 473 

\bibitem[1987]{andre}
Andr\'e, Ph. 1987, in: Protostars and Molecular Clouds, eds. T.~Montmerle
\& C.~Bertout, p.143

\bibitem[Ardila \& Basri(2000)]{2000ApJ...539..834A} Ardila, D.~R.~\& 
Basri, G.\ 2000, ApJ, 539, 834 

\bibitem[Ardila et al.(2002)]{2002ApJ...566.1100A} Ardila, D.~R., Basri, 
G., Walter, F.~M., Valenti, J.~A., \& Johns-Krull, C.~M.\ 2002a, ApJ, 566, 
1100 

\bibitem[Ardila et al.(2002)]{2002ApJ...567.1013A} Ardila, D.~R., Basri, 
G., Walter, F.~M., Valenti, J.~A., \& Johns-Krull, C.~M.\ 2002b, ApJ, 567, 
1013 

\bibitem[Armitage \& Clarke(1996)]{1996MNRAS.280..458A} Armitage, P.~J.~\& 
Clarke, C.~J.\ 1996, MNRAS, 280, 458 


\bibitem[Bardou \& Heyvaerts(1996)]{1996A&A...307.1009B} Bardou, A.~\& 
Heyvaerts, J.\ 1996, A\&A, 307, 1009 

\bibitem[Basri \& Bertout(1989)]{1989ApJ...341..340B} Basri, G.~\& Bertout, 
C.\ 1989, ApJ, 341, 340 

\bibitem[Batalha et al.(2002)]{2002ApJ...580..343B} Batalha, C., Batalha, 
N.~M., Alencar, S.~H.~P., Lopes, D.~F., \& Duarte, E.~S.\ 2002, ApJ, 580, 
343 


\bibitem[Beristain, Edwards, \& Kwan(2001)]{2001ApJ...551.1037B} Beristain, 
G., Edwards, S., \& Kwan, J.\ 2001, ApJ, 551, 1037 

\bibitem[Bertout(2000)]{2000A&A...363..984B} Bertout, C.\ 2000, A\&A, 363, 
984 

\bibitem[Bertout, Basri, \& Bouvier(1988)]{1988ApJ...330..350B} Bertout, 
C., Basri, G., \& Bouvier, J.\ 1988, ApJ, 330, 350 

\bibitem[Bouvier \& Bertout(1989)]{1989A&A...211...99B} Bouvier, J.~\& 
Bertout, C.\ 1989, A\&A, 211, 99 

\bibitem[Bouvier et al.(1993)]{1993A&A...272..176B} Bouvier, J., Cabrit, 
S., Fernandez, M., Martin, E.~L., \& Matthews, J.~M.\ 1993, A\&A, 272, 176 

\bibitem[1999]{bou99} 
Bouvier, J., Chelli, A., Allain, S., et al. 1999, A\&A, 349, 619  (B99)

\bibitem[2003]{bou03}
Bouvier, J., Grankin, K., Alencar, S., et al. 2003, A\&A, in press


\bibitem[Calvet \& Gullbring(1998)]{1998ApJ...509..802C} Calvet, N.~\& 
Gullbring, E.\ 1998, ApJ, 509, 802 

\bibitem[Camenzind(1990)]{1990RvMA....3..234C} Camenzind, M.\ 1990, Reviews 
of Modern Astronomy, 3, 234 

\bibitem[Carpenter, Hillenbrand, \& Skrutskie(2001)]{2001AJ....121.3160C} 
Carpenter, J.~M., Hillenbrand, L.~A., \& Skrutskie, M.~F.\ 2001, AJ, 121, 
3160 

\bibitem[Chelli et al.(1999)]{1999A&A...345L...9C} Chelli, A., Carrasco, 
L., M{\' u}jica, R., Recillas, E., \& Bouvier, J.\ 1999, A\&A, 345, L9 

\bibitem[Collier Cameron \& Campbell(1993)]{1993A&A...274..309C} Collier 
Cameron, A.~\& Campbell, C.~G.\ 1993, A\&A, 274, 309 


\bibitem[2003]{dewarf03}
DeWarf,  L.E., Sepinsky, J.F., Guinan, E.F., Ribas, I., Nadalin, I. 2003,
ApJ, in press


\bibitem[Edwards et al.(1993)]{1993AJ....106..372E} Edwards, S.~et al.\ 
1993, AJ, 106, 372 

\bibitem[1994]{edw94} 
Edwards, S., Hartigan, P., Ghandour, L., \& Andrulis, C. 1994, AJ, 108, 1056

\bibitem[Eiroa et al.(2002)]{2002A&A...384.1038E} Eiroa, C.~et al.\ 2002, 
A\&A, 384, 1038 

\bibitem[Ferreira(1997)]{1997A&A...319..340F} Ferreira, J.\ 1997, A\&A, 
319, 340 

\bibitem[Ghosh \& Lamb(1979)]{1979ApJ...234..296G} Ghosh, P.~\& Lamb, 
F.~K.\ 1979, ApJ, 234, 296 

\bibitem[Gomez de Castro \& Lamzin(1999)]{1999MNRAS.304L..41G} Gomez de
  Castro, A.~I.~\& Lamzin, S.~A.\ 1999, MNRAS, 304, L41


\bibitem[Goodson, Winglee, \& Boehm(1997)]{1997ApJ...489..199G} Goodson, 
A.~P., Winglee, R.~M., \& Boehm, K.\ 1997, ApJ, 489, 199 

\bibitem[Guenther, Lehmann, Emerson, \& Staude(1999)]{1999A&A...341..768G} 
Guenther, E.~W., Lehmann, H., Emerson, J.~P., \& Staude, J.\ 1999, A\&A, 
341, 768 

\bibitem[Gullbring, Hartmann, Briceno, \&
  Calvet(1998)]{1998ApJ...492..323G} Gullbring, E., Hartmann, L., Briceno,
  C., \& Calvet, N.\ 1998, ApJ, 492, 323

\bibitem[Folha & Emerson(1999)]{1999A&A...352..517F} Folha, D.~F.~M.~\& 
Emerson, J.~P.\ 1999, A\&A, 352, 517 

\bibitem[Folha & Emerson(2001)]{2001A&A...365...90F} Folha, D.~F.~M.~\& 
Emerson, J.~P.\ 2001, A\&A, 365, 90 

\bibitem[Hartigan, Edwards, \& Ghandour(1995)]{1995ApJ...452..736H} 
Hartigan, P., Edwards, S., \& Ghandour, L.\ 1995, ApJ, 452, 736 

\bibitem[Hartmann, Hewett, \& Calvet(1994)]{1994ApJ...426..669H} Hartmann, 
L., Hewett, R., \& Calvet, N.\ 1994, ApJ, 426, 669 

%

\bibitem[Hayashi, Shibata, \& Matsumoto(1996)]{1996ApJ...468L..37H} 
Hayashi, M.~R., Shibata, K., \& Matsumoto, R.\ 1996, ApJl, 468, L37 

\bibitem[Herbst, Rhode, Hillenbrand, \& Curran(2000)]{2000AJ....119..261H} 
Herbst, W., Rhode, K.~L., Hillenbrand, L.~A., \& Curran, G.\ 2000, AJ, 
119, 261 

%

\bibitem[Johns \& Basri(1995)]{1995ApJ...449..341J} 
Johns, C.~M.~\& Basri, G.\ 1995a, ApJ, 449, 341


\bibitem[Johns-Krull \& Gafford(2002)]{2002ApJ...573..685J} Johns-Krull, 
C.~M.~\& Gafford, A.~D.\ 2002, ApJ, 573, 685 

\bibitem[Johns-Krull \& Hatzes(1997)]{1997ApJ...487..896J} Johns-Krull, 
C.~M.~\& Hatzes, A.~P.\ 1997, ApJ, 487, 896 

\bibitem[Johns-Krull \& Valenti(2001)]{2001ApJ...561.1060J} Johns-Krull, 
C.~M.~\& Valenti, J.~A.\ 2001, ApJ, 561, 1060 

\bibitem[Johns-Krull \& Valenti(2003)]{preprint} Johns-Krull, C.~M.~\&
  Valenti, J.~A.\ 2003, ApJ, in press

\bibitem[Johns-Krull \& Valenti(2000)]{2000scac.conf..371J} Johns-Krull, 
C.~M.~\& Valenti, J.~A.\ 2000, ASP Conf.~Ser.~198: Stellar Clusters and 
Associations: Convection, Rotation, and Dynamos, 371 

\bibitem[Johns-Krull, Valenti, \& Koresko(1999)]{1999ApJ...516..900J} 
Johns-Krull, C.~M., Valenti, J.~A., \& Koresko, C.\ 1999, ApJ, 516, 900 

\bibitem[Johns-Krull, Valenti, Saar, \& Hatzes(2001)]{2001csss...11..521J} 
Johns-Krull, C.~M., Valenti, J.~A., Saar, S.~H., \& Hatzes, A.~P.\ 2001, 
ASP Conf.~Ser.~223: 11th Cambridge Workshop on Cool Stars, Stellar Systems 
and the Sun, 11, 521 

%
%
%

\bibitem[Koenigl(1991)]{1991ApJ...370L..39K} Koenigl, A.\ 1991, ApJl, 370, 
L39 

\bibitem[Lai(1999)]{1999ApJ...524.1030L} Lai, D.\ 1999, ApJ, 524, 1030 

\bibitem[Lamzin(1995)]{1995A&A...295L..20L} Lamzin, S.~A.\ 1995, A\&A, 295, 
L20 

\bibitem[Lin, Bodenheimer, \& Richardson(1996)]{1996Natur.380..606L} Lin, 
D.~N.~C., Bodenheimer, P., \& Richardson, D.~C.\ 1996, Nature, 380, 606 

\bibitem[Lopez-Martin(2003)]{Lopez-Martin2003} 
Lopez-Martin, L., Cabrit, S., Dougados, C. 2003, A\&A, in press

\bibitem[Martin(1996)]{1996ApJ...470..537M} Martin, S.~C.\ 1996, ApJ, 470, 
537 

\bibitem[M{\' e}nard \& Bertout(1999)]{1999osps.conf..341M} M{\' e}nard, 
F.~\& Bertout, C.\ 1999, NATO ASIC Proc.~540: The Origin of Stars and 
Planetary Systems, 341 


\bibitem[Meyer, Calvet, \& Hillenbrand(1997)]{1997AJ....114..288M} Meyer, 
M.~R., Calvet, N., \& Hillenbrand, L.~A.\ 1997, AJ, 114, 288 


\bibitem[Montmerle, Koch-Miramond, Falgarone, \& 
Grindlay(1983)]{1983ApJ...269..182M} Montmerle, T., Koch-Miramond, L., 
Falgarone, E., \& Grindlay, J.~E.\ 1983, ApJ, 269, 182 


\bibitem[Montmerle, Grosso, Tsuboi, \& Koyama(2000)]{2000ApJ...532.1097M} 
Montmerle, T., Grosso, N., Tsuboi, Y., \& Koyama, K.\ 2000, ApJ, 532, 1097 

%

\bibitem[2001]{muz01}
Muzerolle, J., Hartmann, L., \& Calvet, N. 2001, ApJ, 550, 944


\bibitem[Oliveira, Foing, van Loon, \& Unruh(2000)]{2000A&A...362..615O} 
Oliveira, J.~M., Foing, B.~H., van Loon, J.~T., \& Unruh, Y.~C.\ 2000, 
A\&A, 362, 615 


\bibitem[Petrov et al.(1996)]{1996A&A...314..821P} Petrov, P.~P., 
Gullbring, E., Ilyin, I., Gahm, G.~F., Tuominen, I., Hackman, T., \& Loden, 
K.\ 1996, A\&A, 314, 821 

\bibitem[Petrov et al.(2001)]{2001A&A...369..993P} Petrov, P.~P., Gahm, 
G.~F., Gameiro, J.~F., Duemmler, R., Ilyin, I.~V., Laakkonen, T., Lago, 
M.~T.~V.~T., \& Tuominen, I.\ 2001, A\&A, 369, 993 

%

\bibitem[Rebull, Wolff, Strom, \& Makidon(2002)]{2002AJ....124..546R} 
Rebull, L.~M., Wolff, S.~C., Strom, S.~E., \& Makidon, R.~B.\ 2002, AJ, 
124, 546 

\bibitem[Romanova, Ustyugova, Koldoba, \& 
Lovelace(2002)]{2002ApJ...578..420R} Romanova, M.~M., Ustyugova, G.~V., 
Koldoba, A.~V., \& Lovelace, R.~V.~E.\ 2002, ApJ, 578, 420 

\bibitem[Romanova, Ustyugova, Koldoba, Vick,
  Lovelace(2003)]{2003ApJ...inpress} Romanova, M.~M., Ustyugova, G.~V.,
  Koldoba, A.~V., Vick J.W., ,\& Lovelace, R.~V.~E.\ 2003, ApJ, in press

%
%

\bibitem[Shu et al.(1994)]{1994ApJ...429..781S} Shu, F., Najita, J., 
Ostriker, E., Wilkin, F., Ruden, S., \& Lizano, S.\ 1994, ApJ, 429, 781 

\bibitem[Stassun, Mathieu, Mazeh, \& Vrba(1999)]{1999AJ....117.2941S} 
Stassun, K.~G., Mathieu, R.~D., Mazeh, T., \& Vrba, F.~J.\ 1999, AJ, 117, 
2941 

\bibitem[Smith et al, 2003]{2003AAinpress} Smith, K., Pestalozzi, M.,
G\"udel, M., Conway, J., Benz, A.O. 2003, AA, in press
(astro-ph/0305543)

%

\bibitem[Terquem \& Papaloizou(2000)]{2000A&A...360.1031T} Terquem, C.~\& 
Papaloizou, J.~C.~B.\ 2000, A\&A, 360, 1031 

%
%

\bibitem[Unruh, Collier Cameron, \& Guenther(1998)]{1998MNRAS.295..781U} 
Unruh, Y.~C., Collier Cameron, A., \& Guenther, E.\ 1998, MNRAS, 295, 781 

%
%

\bibitem[Vrba, Chugainov, Weaver, \& Stauffer(1993)]{1993AJ....106.1608V} 
Vrba, F.~J., Chugainov, P.~F., Weaver, W.~B., \& Stauffer, J.~S.\ 1993, 
AJ, 106, 1608 

\bibitem[Woitas et al.(2002)]{2002ApJ...580..336W} Woitas, J., Ray, T.~P., 
Bacciotti, F., Davis, C.~J., \& Eisl{\" o}ffel, J.\ 2002, ApJ, 580, 336 

%
\end{chapthebibliography}
\end{document}